%% file: arxiv.tex
\documentclass[conference]{IEEEtran}
\IEEEoverridecommandlockouts
\usepackage{cite}
\usepackage{amsmath,amssymb,amsfonts}
\usepackage{algorithmic}
\usepackage{graphicx}
\usepackage{textcomp}
\usepackage{xcolor}

\usepackage{subfigure}
\usepackage[english]{babel}
\usepackage[utf8x]{inputenc}
\usepackage{caption}
\usepackage[normalem]{ulem}
\usepackage{array}
\usepackage{multirow}
\usepackage{booktabs}

\newcolumntype{R}[1]{>{\raggedleft\arraybackslash }m{#1}}
\newcolumntype{L}[1]{>{\raggedright\arraybackslash }m{#1}}
\newcolumntype{C}[1]{>{\centering\arraybackslash }m{#1}}

\def\BibTeX{{\rm B\kern-.05em{\sc i\kern-.025em b}\kern-.08em
    T\kern-.1667em\lower.7ex\hbox{E}\kern-.125emX}}
\begin{document}

\title{Exploring the Effectiveness of Face-to-face Mixed Reality for Teaching with Chalktalk
}

\author{
 \IEEEauthorblockN{1\textsuperscript{st} Zhenyi He}
 \IEEEauthorblockA{\textit{New York University} \\
 zh719@nyu.edu}
 \and
 \IEEEauthorblockN{2\textsuperscript{rd} Ken Perlin}
 \IEEEauthorblockA{\textit{New York University} \\
 ken.perlin@gmail.com}

}

\maketitle

\begin{abstract}
Teaching that uses projected presentation media such as slide-shows lacks support for dynamic content whose form and behaviors require live changes during a lecture. Recent software alternatives such as the Chalktalk software platform allow creation of interactive simulations in arbitrary sequences and combinations within presentations. These more dynamic solutions, however, do not optimize for face-to-face interactions: eye-contact, gaze direction, and concurrent awareness of another person's movements together with the presented content. To explore the extent to which these face-to-face interactions may improve learning and engagement during a lecture, we propose a Mixed Reality (MR) platform that places Chalktalk's behaviors and simulations within a mirrored virtual world environment designed for face-to-face, one-on-one interactions. We compare our system with projected Chalktalk to evaluate its relative effectiveness for learning, retention, and level of engagement.
\end{abstract}

\input{body}

\bibliographystyle{IEEEtran}
\bibliography{sample}


\end{document}

%% file: body.tex
\section{Introduction}

Projected presentation media continue to be standard tools used for teaching. The most common and familiar software solutions, including Powerpoint and Sharepoint, are modeled after the slide-projector show, and provide a storyboard-like format, allowing for sequential text, images, and pre-defined animation sequences to be viewed one step at a time. Other options such as Prezi~\cite{perron2010review} additionally allow for branching sequences. These media usually supplement an oral presentation, which the speaker can perform with some flexibility for improvisation. Since traditional slideshows are sequential, they might discourage the speaker from conducting a non-sequential, more flexible presentation.
If members of the audience ask questions whose answers don't lie in the slides, then the lecturer might need to abandon the slide presentation and rely entirely on speech, hand gestures, and sometimes reference objects. This becomes particularly problematic when the concepts under discussion possess dynamic behaviors or require live modifications and interactions that cannot be anticipated before the presentation is prepared. 

Designed in-part to overcome these issues, a recent open-source platform called Chalktalk ~\cite{perlin2016future, nunes2017atypical, perlin2018chalktalk, Perlin2018-by} supports not only arbitrary sequencing of content, but also the creation of interactive simulations that can be combined arbitrarily as well. However, the result is still lacking in terms of presenter-audience engagement due to the way the presenter, audience and projected presentation are located and oriented. The typical triangular format--where the presenter stands in front or to the side of the projected content towards an onlooking audience--is suboptimal. It may lead the presenter to block the view of the content in many cases, and it requires the presenter and audience to divide their attention between each other and the presented content. An alternative format would facilitate face-to-face interactions and help participants to better focus on the presenter's movements and gaze, as well as the content, with less need for attention switching.

In previous research, systems have been implemented using face-to-face interaction to improve collaboration~\cite{ishii1992clearboard, harrison1995transparent, clearboard1993}, communication~\cite{mmspace16} and other interactions~\cite{transwall14}, and recorded lectures. Simulation of face-to-face interactions is also used in teleprompting systems, which allow a speaker to gaze at the camera naturally while reading from a projected script. Our particular focus has been to investigate the effectiveness and configuration of face-to-face interactions with respect to learning. To explore the extent to which these interactions may improve learning and engagement in the context of a lecture, we have developed a Mixed Reality (MR) platform designed for learning in a face-to-face environment. We conducted a user study in which we used projected Chalktalk and our MR system to present equivalent lessons on matrix transformations for computer graphics. To summarize, our contribution is to provide (1) a protocol to render information from a behavior server such as Chalktalk into a run-time 3D engine that supports VR such as Unity, (2) observations from our comparison between projected Chalktalk and our MR platform, and (3) guidelines for immersive presentation design.

\section{Related Work}

ClearBoard ~\cite{ishii1992clearboard, harrison1995transparent, clearboard1993} creates a shared workspace in which two users can collaborate with each other remotely without losing the advantages of in-person face-to-face interactions--namely eye contact, gaze, and gestures. This is achieved using a transparent glass window on which remote collaborators can draw (as they would on a drafting table), and through which the collaborators can see each other in the correct--non-reversed--orientation. In short, ClearBoard preserves users'  ``gaze awareness," which means that a user can tell where his or her collaborator is looking on the screen. Thus, conversations about content drawn on ClearBoard approximate conversations around a whiteboard. The ClearBoard implementation has underwent a number of iterations, where version 1 uses markers and version 2 provides software to enable drawing with digital pens, as well as improvements to allow for video recordings.

The goal of MMspace~\cite{mmspace16, mmspace17, mmspace18} is to allow for face-to-face social interactions and telepresence in the context of small group remote conferences. To achieve this, the system uses custom-built mechanical displays on which images of remote participants are projected, and which react to users' movements. These are called ``kinetic avatars." Using data recognized from users' head movements and gestures, the system reorients the avatar displays to emulate human behaviors such as head gestures and gaze awareness. Pairs of participants can maintain eye contact with each other and remain aware of each other's focus. (Users sit at a small table facing the displays.) The movements of the avatars emulate human gestures such as nodding or changing gaze direction.

Fuhrmann et al. ~\cite{3dslides} propose a framework for presentations in virtual reality (VR), extending the 2D content of projected slides to support ``applets" that can be embedded into the 3D region of a ``slide" in the proposed system. During the presentation, simpler widgets can be instantiated to show simple concepts or act as controls (e.g. a sliders) for input to the applets. The 3D content is placed within the context of a linear presentation sequence. The system is also meant to work with multiple users in various configurations. For example, the presenter might use a head-mounted display (HMD) while the audience uses polarized glasses to view the content, or all participants might use HMDs.

Tan et al.~\cite{gazeawareness} identify two main problems with traditional projected presentations: (1) for very large audiences, the presenter must often use a large display for all to see, in which case the display is out-of-reach of the presenter, who cannot use gestures and gaze to supplement the presentation; (2) it can be challenging for the audience to figure where on the screen the presenter is looking when the presenter stands at an angle to the display and audience. To remedy this problem, the authors develop a projected system: the presenter and audience stand/sit in front of their own projection screens in separate locations while live video feed from both sites records the presenter and audience. On the presenter side, the projection shows the audience feed. On the audience side, the projection shows an alpha-blended (roughly speaking) composition of the presenter's video feed and the presentation content. This way, the audience can focus on the presenter, who becomes part of the content.

Transwall~\cite{transwall14} is a double-sided transparent display built for face-to-face interaction between a pair of co-located users in physical reality. It is implemented with holographic screen film and video projections. (The authors suggests that the system would have benefited newer technology such as certain transparent OLED screens.) The system reverses the image on both sides. In building the system, the authors identified various interaction design factors: the users should be able to touch the same board on the board from either side without interference, interaction options should be configurable depending on the side of the board the touch interaction takes place, images should be flippable between sides of the display, the real world should overlap with the virtual contents of the display (so the physical background can be incorporated into the graphics of software running on the display), and the level of transparency should be controllable. To explore these design factors, the creators developed several mini-games for touch interactions and face-to-face communication. For example, a word-finding game ``Flip and spell" has users use touch controls to flip text from one side of the board to their own to form words and earn points.

GAZE groupware~\cite{gaze1999} is a networked video conference system designed to showcase how gaze direction (1) affects our perception of who is speaking and (2) helps us track the subject of the conversation at-hand. The GAZE software employs eye-tracking hardware to determine the position of remote users' gaze and uses this information to set the orientation of avatars (live video feeds of the users' heads) in a virtual round-table environment displayed on a computer monitor in front of the user. The eye-tracking provides visual feedback in the form of colored lights and allows the user to select and examine shared virtual documents. The associated study suggests that ``conveying gaze direction - especially gaze at the facial region - eases turn taking, allowing more speaker turns and more effective use of deictic verbal references" with the caveat that this conclusion does not imply improved performance for all collaborative tasks.

cAR/PE!~\cite{carpe03} is another group video conferencing system for remote collaboration and communication that seeks to emulate face-to-face interaction. Each remote user is represented as an avatar (head recording) that is placed in a shared augmented reality (AR) scene to be displayed locally on a computer monitor. The user controls movement of the avatar by using a mouse within a custom VR/AR framework, while the system tracks the user's head movement, which adjusts the angle of the avatar's display. Users can interact with and discuss 3D geometric objects.

The Multiview~\cite{multiview} conference system focuses on creating spatial faithfulness between users's environments over a network to facilitate face-to-face conversations. The authors define spatial faithfulness as the ``ability to preserve spatial relationships between people and objects." In other words, the system should avoid distorting the perspective and apparent position and orientation of remote users. Two groups of people sit at opposite ends of a table, where all participants on one side see the other side clearly. Multiview uses a multi-camera system to display different views of other users to a given user depending on the position and perspective of users within a ``viewing zone." Multiview helps preserve gaze, allowing for deictic cues and gestures.

Inspired by Clearboard, Clearspace~\cite{clearspace2011} uses the mirrored world paradigm (flipping on x) to facilitate multi-user collaboration in a shared virtual reality space (displayed to the user on a monitor). It uses head and hand tracking. It incorporates portals into the design to allow users to move between sides and views of the remote users' boards while keeping spatial/physical consistency by transforming (scaling) the users' avatars when necessary--for example when different users access the software from differently-sized devices such as a laptops and mobile phones, which might skew their relative scales. Portals also lead to virtual documents (e.g. drawings, pdfs, and others).

\section{Introduction to Chalktalk}

\begin{figure}[h]
\centering
\includegraphics[width=0.3\textwidth]{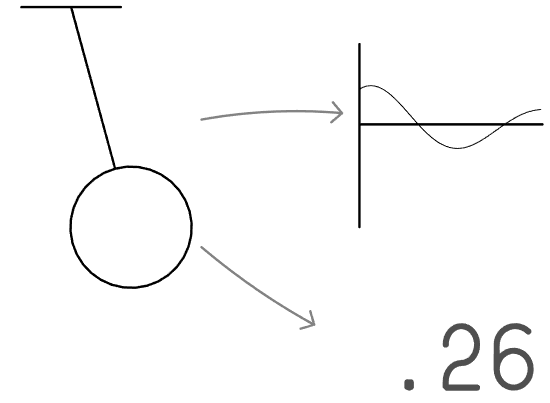}
\caption{left: the user has swung a pendulum sketch using a mouse gesture; bottom: a link from the pendulum to the Chalktalk draw canvas displays the data output of the pendulum, which is a floating point number representing the pendulum's angle in radians; right: an xy plot sketch receives the same data from the pendulum and uses it to display a curve representing the motion of the pendulum over time. Since the pendulum outputs a float, sketches to which the pendulum is connected receive the value and are free to interpret it according to the logic defined in the sketch. This means that sketches that expect a float are compatible with each other, and so multiple sketches can be linked and composited together to form larger systems.}
\label{fig:pendulum_sys}
\end{figure}

Chalktalk is a web browser-based 3D presentation and communication tool that provides a blackboard-like interface in which the user's drawings are recognized as animated and interactive 3D (or 2D) sketches. Each sketch is programmed to demonstrate a different set of ideas and concepts. Sketches can be linked together logically to generate increasingly complex behaviors during a live presentation ~\ref{fig:pendulum_sys}. Chalktalk contains an extensible library of sketches, so the user can add new sketches by programming them in JavaScript, either before or during the presentation, using a built-in code editor. In addition, Chalktalk allows for slideshow-like text, embedded images, and free-hand drawings. Sketches can be panned off-screen when additional space is needed, thereby making erasure less frequent. In short, Chalktalk offers traditional presentation features coupled with interactive and programmatic content.

There are several reasons that we chose Chalktalk for our behavior server (which would provide behaviors and user interaction functionality for our system). First, it is open-source software, which means it can be integrated into different platforms, and data can be collected and transferred freely throughout systems that interact with it. Chalktalk's original architecture follows the client/server model. The server (implemented in Node.js) runs in the back-end to maintain all browser clients, and it communicates with clients by using a websocket. Based on this design, we can add another node to communicate with the Chalktalk server so that the websocket can send data from other nodes to the clients, and also so that we can collect the data we need from the client side. Second, Chalktalk's draw space is an infinite canvas rather than a linear sequence. Because our system is an MR platform, the presentation view area is potentially the entire world environment. Thus, our system could benefit from Chalktalk's open draw-space. Third, Chalktalk is a 3D presentation tool. Although the input method and visuals for Chalktalk are projected to 2D, all content in Chalktalk window is initially 3D. In our MR system, this content need not be projected to a 2D screen. It can instead be viewed directly as 3D objects within the world. 3D content support is the key to creating an immersive environment in which concepts can be visualized and understood intuitively. This is crucial for nearly any subject ranging from geometry to procedural animation. The aforementioned features and drawing capabilities made Chalktalk the ideal behavior server for our system.

\section{Design Principles and Implementation Details}
\subsection{System Overview}

Our system supports not only VR headsets but also AR devices, since our system does not require much physical space. The experience can be run anywhere -- at home, at the office or in a park. We chose the Oculus Rift as the VR headset for easy setup, and the Google Pixel phone as the AR device for an outdoor scenario. We use Unity as our development platform, because it can readily accommodate executables for different hardware, with no changes required to scene design. Figure~\ref{fig:phyPlace} shows the physical location of participants and the presenter for different devices. Here we assume that the Oculus sensors are placed on a table in the front. We have users stand aside the table facing parallel to the side of the table. The users can walk around the content, which is displayed on a plane perpendicular to the table. Figure~\ref{fig:virPlace} shows the immersive environment for both the presenter and the audience. The presenter and the audience are located at opposite sides of the presentation content, looking at each other face-to-face. We achieve this with a virtual mirror effect, which we describe in more detail in the following section. Here we define both presenter and audience as one category of role. We render remote users with different roles in a mirrored way according to the content. In the following section we discuss our rationale and implementation for this face-to-face configuration.

\begin{figure}[h]
\centering
        \includegraphics[width=0.45\textwidth]{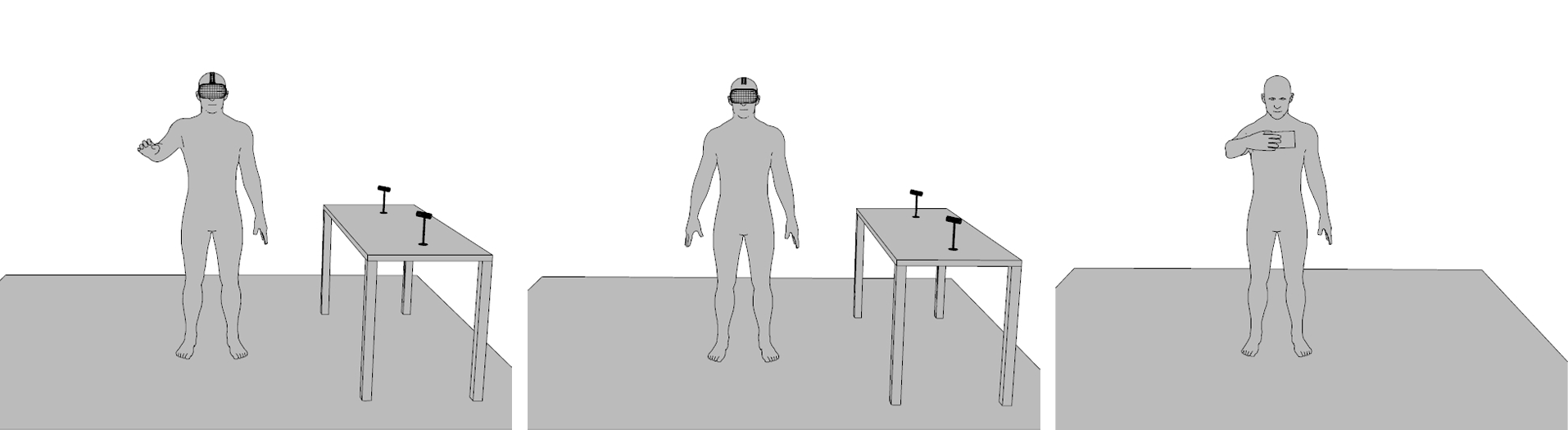}
        \caption{Physical configuration for participants and devices. Left is for presenter with Oculus Rift; middle is for audience with Oculus Rift and right is for audience with AR phone.}
        \label{fig:phyPlace}
\label{fig:phyPlace}
\end{figure}

\begin{figure}[h]
\centering
        \includegraphics[width=0.45\textwidth]{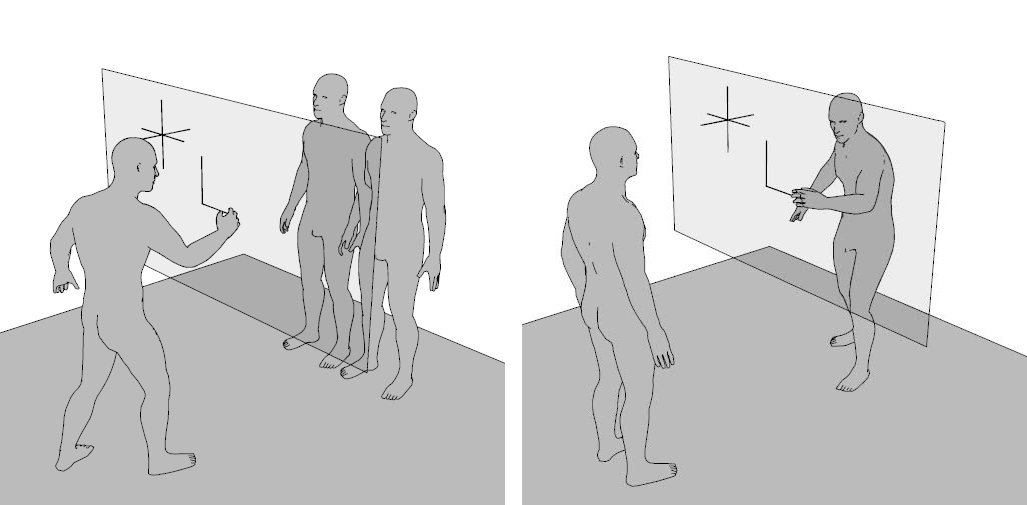}
        \caption{MR view for presenter and audience. Left is the presenter view. Presenter can see all audience; Right is the audience view. And only presenter is visible to him.}
        \label{fig:virPlace}
\label{fig:virPlace}
\end{figure}

\subsection{Face-to-Face Presentation}\label{sec:a}
Two main approaches to presentation are used widely: (1) the blackboard/whiteboard and (2) the projected slideshow.  In the case of blackboard/whiteboard presentations, the audience usually faces the board directly, and the presenter stands at or near the board at all times. Standing at an angle to the board and audience, the presenter cannot simultaneously focus on drawing the presentation content and maintaining eye-contact with the audience ~\cite{gazeawareness}. As a result, the presenter must change focus repeatedly, and the audience, too, cannot maintain eye contact with the presenter when looking at content on the board. There always exists a context switch to and from the board that interrupts interaction between presenter and audience. Sometimes the presenter might also obstruct the board ~\cite{3dslides} on at least one side while drawing. The multiple cameras or displays that conferences such as TED ~\cite{gazeawareness} afford are unlikely to be available for the typical classroom setting, and they must be configured ahead of time. For the classroom setting, it is possible for the audience to become lost when the presenter describes something being drawn while blocking that drawing from view. By the time the view of this new drawing becomes clear, some in the audience might have lost track of the argument. 

For the projected slideshow case, the presenter, audience, and board typically have the same spatial configuration as in the blackboard/whiteboard case, with the difference that slideshow content is created and sequenced prior to the presentation, and the presenter is less likely to physically block the board now that the content need not be drawn. Predefined content might save time and reduce the number of context switches for the presenter, but the presentation is fixed to that predefined content. If clarification becomes necessary during a lecture, the presenter must fall-back to drawing on a whiteboard or using spoken word descriptions. To follow the lecture, the audience must still alternate its focus between presenter and content or choose not to pay attention to the presenter, thereby reducing emotional engagement.

We found that face-to-face interactions are important to the audience ~\cite{ishii1992clearboard, gazeawareness, gaze1999}, so we designed our MR system with that in mind. We render Chalktalk's sketches to an infinite transparent board, (we will call this the "MR board"). If we were to simply place each user across from one another with the MR board between them, one user would see reversed content. To avoid this situation, we follow the convention of ~\cite{ishii1992clearboard}, whereby each client faces the other in left-right mirror reversal as though looking into a mirror. In this way, gaze direction is preserved so that the presenter and audience can establish eye-contact while appearing to look at the same objects with text non-reversed for both. Since the transparent MR board lies between presenter and audience, the content itself is not blocked.

We render Chalktalk's sketches to a shared infinite transparent board, (we will call this the "MR board"). We found that face-to-face interactions are important to the audience ~\cite{ishii1992clearboard, gazeawareness, gaze1999}.  Yet if two people were to stand physically across from one another with a pane of glass between them, text on that pane of glass would appear backwards for one of them.  Therefore, inspired by previous work on "Clearboard," we mirror the presenter's view of students, and vice versa, such that each appears to the other left-right mirror reversed on opposite sides of the MR board. In this way, gaze direction is preserved, allowing presenter and audience can establish eye-contact while appearing to look at the same objects with text non-reversed for both. Since the transparent MR board lies between presenter and audience, the content itself is not blocked.

\subsection{One-to-One Experience}
We also wanted to explore how immersive one-on-one interactions might make presentations more engaging and effective for learning. In this experience, the teacher should be aware of all students, but each student should be aware of only the teacher. Thus, we built our system to allow for a multi-user presentation scenario in which one presenter addresses multiple (remote or local) audience members at the same time, while audience members are shown only the teacher. To clarify the implementation, only the presenter's avatar is rendered for each audience member. The presenter, however, sees all audience members' avatars rendered, and audience members can provide visual feedback to the presenter via body language (We send each audience member's local transform information to the presenter over a network.) All participants can also hear the same audio and communicate over a group call. This means that although audience members cannot see each other, they can still interact verbally--for example, when asking the presenter questions that everyone might want answered.

\subsection{Holojam Network Framework}
\begin{figure}[h]
\centering
\includegraphics[width=0.3\textwidth]{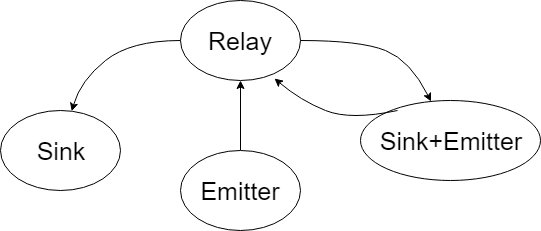}
\caption{Holojam framework (relay, sink, and emitter)}
\label{fig:holojam}
\end{figure}

Our system provides a multi-user experience during the presentation. To communicate across clients, we opted to use Holojam~\cite{perlin2016future} as the network framework for data transmission, as it supports multiple clients, including the Unity game engine that renders on the client side and Node.JS, which runs Chalktalk on the server side. Holojam is designed as an untethered VR headset system that enables multiplayer shared-space experiences. It is an open-source tool that has been used for several projects~\cite{velho2017situated,masson2017holo,he2017physhare}. We customize its data protocol for our MR environment. As shown in figure~\ref{fig:holojam}, Holojam provides many components to handle communication between nodes. Each node is assigned at least one role: relay, emitter, and/or sink. A relay is the central node of Holojam designed to broadcast and receive all the data from other nodes. There can be only one relay per Holojam network. An emitter sends its own data to the relay with a unique label. A sink receives and processes the information associated with specific labels every frame. Thus, different nodes can talk to each other using labels. Holojam's data format is sufficiently flexible for our system. It supports vector3, vector4, byte, and other data types, so it is easy to synchronize the position and orientation data we must send between clients. Because Holojam supports raw byte data, customized encoding and decoding of all kinds of data is possible. 

\begin{figure}[h]
\centering
\includegraphics[width=0.45\textwidth]{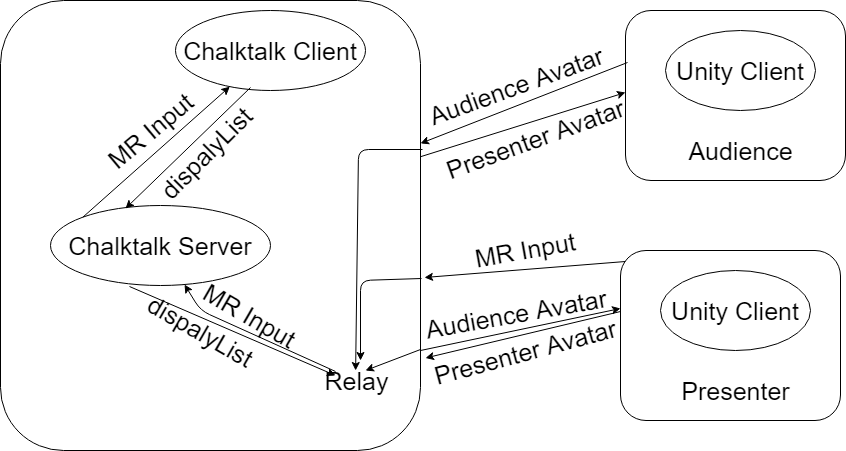}
\caption{We need another architecture figure to explain the data transfer flow}
\label{fig:architecture}
\end{figure}

For the MR system's user input(see figure~\ref{fig:architecture}), the game engine sends input data to the behavior server through the Holojam relay. After the behavior server processes the data, rendering information is sent back to the engine over the the relay. To synchronize the visualization of the user, we record details about each player at every frame and assign different labels to differentiate players from each other. As a result, our system supports multiple users that can see each other, and is able to render draw data received from the behavior server with no knowledge of the server's own rendering pipeline.

\subsection{Protocol Between Behavior Server and Game Engine} 

After we decided to use Holojam as the network framework, we designed an intermediate data protocol to encode the data from Chalktalk and decode the packets in Unity. This simple data protocol decouples the behaviors/functionalities provided by the behavior server from the game engine's rendering system. This makes it possible to use different engines and behaviors other than the ones we initially chose (Unity and Chalktalk), which makes our system flexible and reusable.



\section{Evaluating through the Matrix Lecture}
\subsection{Experiment Design}
\begin{figure}[h]
  \centering
  \subfigure[Introduction to matrices]{\includegraphics[width=0.48\textwidth]{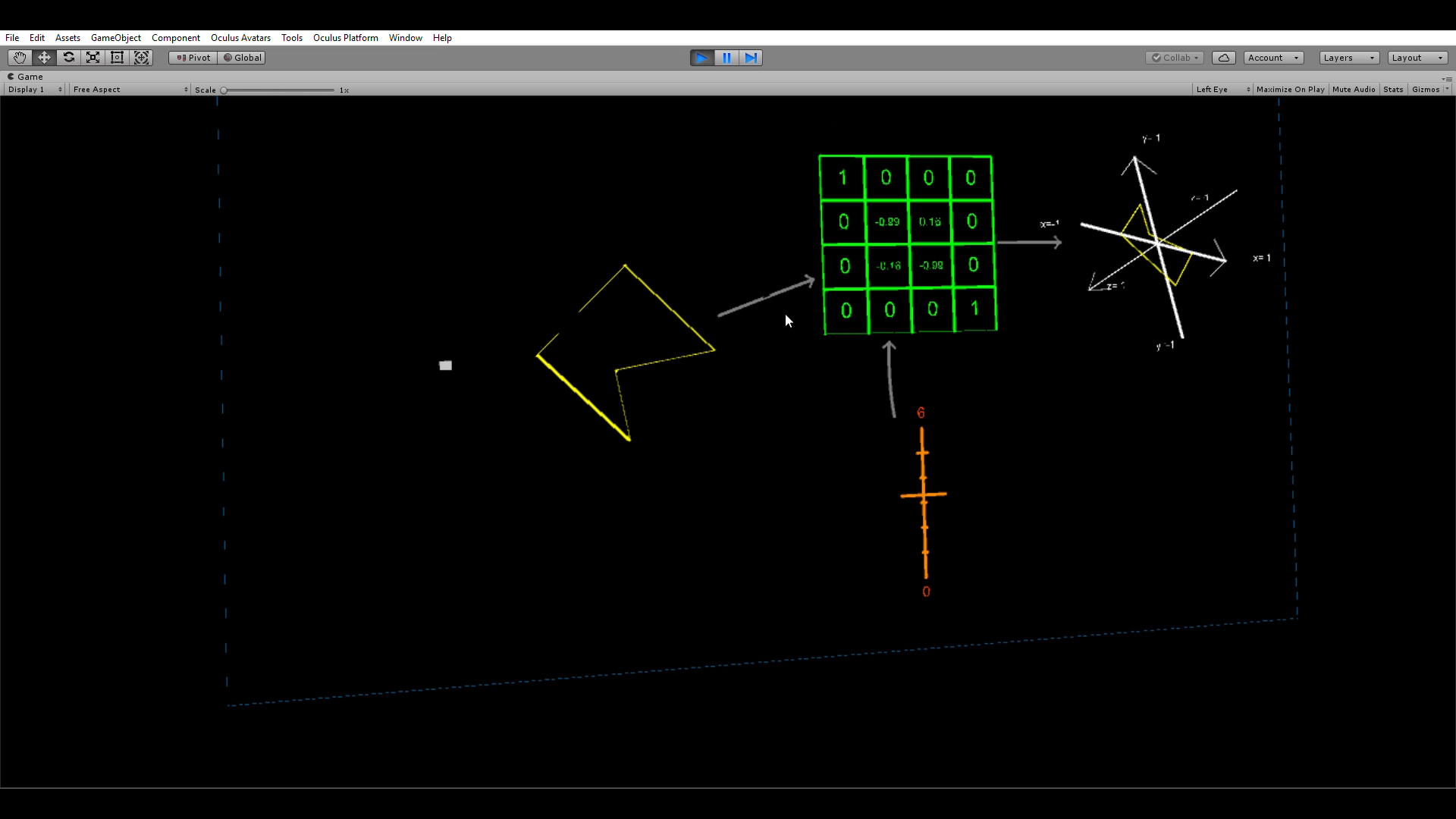}\label{fig:expMatrix:1}}
  \subfigure[Composition of matrix transformations]{\includegraphics[width=0.48\textwidth]{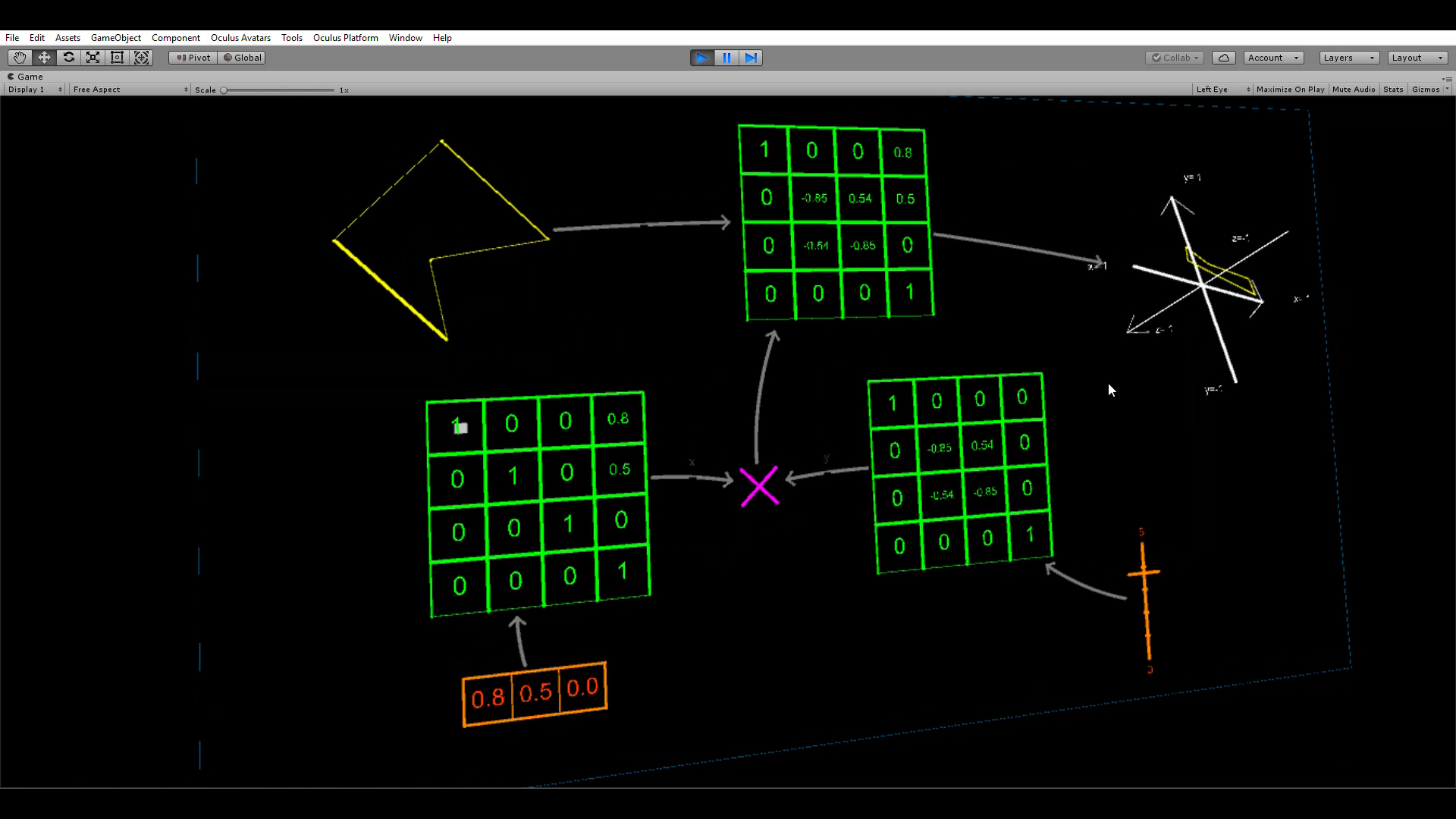}\label{fig:expMatrix:2}}
  \caption{Experiment for part A: introduction to matrices and part B: composition of matrix transformations.}
  \label{fig:expMatrix}
\end{figure}

To evaluate our system on effectiveness for learning, retention, and level of engagement, we chose 4x4 matrix transformations as the presentation topic. We focused on 3D visualization of how matrices apply to objects, for those already familiar with linear algebra and matrix calculations. The content covered in the topic is translation and rotation matrices, followed by a demonstration that matrix multiplication is not commutative. We present this topic To conduct a realistic presentation, we invited a professor who teaches computer graphics to present a lecture on matrix transformation, and recorded the talk as a template. The reason we chose this topic is that (1) it is complicated enough so that the presenter probably could not easily describe it via words only, and (2) it requires dynamic input to show the transformation idea in an intuitive fashion, which is a key feature of Chalktalk we wanted to incorporate, and (3) the entire presentation can be done in under 10 minutes. That is sufficient time for audiences to experience the corresponding platform along with the learning experience. Figure~\ref{fig:expMatrix} shows two parts of the lecture. Figure~\ref{fig:expMatrix:1} shows how transformation matrix is applied to geometry and figure~\ref{fig:expMatrix:2} shows a matrix multiplication operation and how varying the order of matrix multiplication impacts the result. 

To evaluate our system using projected Chalktalk, we included some specific activities during the experiments for both platforms. The presenter in both environments (1) made gestures during the presentation to draw audience's attention, (2) moved the cursor to point at some part of the content to see to what extent the audience was able to follow, (3) used deictic words ~\cite{hickmann2006space} to see to which extent the audience could follow those, and (4) used Chalktalk's pan operation to shift the entire view to learn how shifting content off of the visible projected display area influenced the experience. In contrast, in VR Chalktalk, because it models an ``infinite screen", even when content is panned, all of the content still remains in view.

\subsection{Apparatus and Participants}
We conducted a subject-specific controlled experiment by giving a presentation using both projected Chalktalk and our system with Oculus Rift. We recruited 8 participants (P1-P8) via email and word-of-mouth, including undergraduate and graduate students from a local university studying either Computer Science or Art and Design. The participants are required to have taken a linear algebra class before. The participants (50\% female) are between the ages of 22 and 26 (M=23.71, SD=1.50) and come from various countries. According to the answers to the pre-screening questionnaire, 100\% have tried VR before. All participants had also seen Chalktalk previously. This helps reduce the "novelty effect" of both presentation formats. We are particularly interested in user interview feedback for our study. We refer to this feedback as part of our results.

\subsection{Procedures}
For each experiment, we have two participants enter the MR experience at the same time, standing in different physical locations isolated from one another. One researcher serves as the investigator, in charge of instructing participants in the study and taking notes. Another researcher fulfills the role of the presenter. We set up a video camera and a computer next to the experience space to record all the real-world events and participants. We simultaneously record the screen of the MR Unity clients. Each experiment is composed of two parts, A and B. Part A is the matrix lesson presented in projected Chalktalk and part B is the equivalent lesson in the MR experience in the Oculus Rift. After every pair of experiments, we swap the order of A and B to reduce the potential for bias.. For part A, we place the presenter behind a laptop and to the side of a large monitor displaying the content of the presentation run on the laptop. The audience sits facing the display and presenter, as would students in a large lecture hall. For part B, the presenter and participants stand in the VR spatial configuration described in figure~\ref{fig:phyPlace}. After each part, participants take a short break and then proceed to the second part. Upon the completion of all sessions, participants complete a questionnaire about their experience and then join a semi-structured exit interview. We follow-up the next day with questions about retention (memory longevity). All factors are evaluated using the 7-pt Likert scale.

\subsection{Measurements}
In addition to the questionnaire, our system records gaze information for each user (audience and presenter) in VR. The information includes the intersection point between users' gaze direction and the MR board, and their focus among three users. Focus information is recorded with respect to each person. To each user, when he or she is not looking at other users in VR, we record NONE with its start and end timestamp. When a user looks at one or more users at the same time, we record the user's name with start and end timestamp. When two users are focusing on each other at the same time, it is an eye contact event happened.

\section{Analysis and Observations}

\subsection{Data Analysis}

We conduct repeated measures using the MANOVA model on the quantitative results to analyze our data. We test the homogeneity of variance for each measure, and apply Greenhouse-Geisser correction when it is violated. We perform Bonferroni's test for post-hoc analysis.
All the data analysis was performed using SPSS. The obtained results are displayed with p-values for statistically significant differences, paired with mean values and confidence intervals. 

\subsection{Awareness Results}

First we check the awareness of the presenter in VR and in reality ($M_{VR}$=$6.17$, $SD_{VR}$=$0.083$, $F(2,8)$=$0.172$, $p$>$0.5$). It turns out that awareness of the presenter does not vary greatly when comparing between the experience in VR and in reality. Although each environment has a different spatial configuration (face-to-face in VR, triangular audience/presenter/board relationship in reality) the audience is always aware of the presenter's presence, never "tuning him or her out" completely. In other words, the location of the presenter has very little impact on how aware and focused the audience is on the presenter.

\subsection{Eye Contact Results}

Next, we check the frequency of eye contact between the audience and presenter in VR and reality environments ($M_{VR}$=$2.5$, $M_{reality}$=$6.7$, $F(2,8)$=$35.59$, $p$<$0.01$). The results contradict our initial assumptions from before the experiment: Since the audience and presenter are placed directly opposite each other (with the content in-between) in the VR scenario, we assumed that eye contact would be more frequent and commonplace than in the real environment.
We also check additional data collected during the VR experiment that represents the points in the world where participants were focusing at all times (labeled as audience A, audience B, presenter, and MR board). 2 out of 8 user study participants had greater than 16 instances of eye contact with presenter. However, in the questionnaire they indicated that they perceived less eye contact with presenter in VR (both score 2) than in reality (score 6). From the video recordings of the experiment in the reality environment, we find that only two audience members (P5, F) and (P7,F) shifted their heads when the presenter used hand gestures. During the post-experiment interview, P3(M) and P7(F) mentioned that the design of the avatar did not have eyes. (To clarify, the avatar did have eyes, but these were stylized and did not have realistic eyeballs). This is one main reason that the audience had less eye contact with the presenter in VR than in reality. With the information we collected shown above, we reach some conclusions from the eye-contact results: (1) Our avatar design prevents audience members from perceiving eye-contact, and yet (2) although the audience does not believe the presenter's avatar has eyes, they do look at the presenter's avatar without consciously noticing. (3) In the presentation in reality, the audience thinks that eye contact is more frequent than in VR even when the presenter focuses on the presentation content shown in a private display nearly at all times. The results representing the audience's opinion on the importance of eye contact (according to the post experiment surveys) are $M$=$5.17$, $SD$=$1.47$. The audience does consider eye contact to be an important factor during the presentation, however, they did not engage in it very much during the experiments. P7(F) expressed that eye contact is more important for the presenter to receive feedback from the audience, but is less important to the audience.

\subsection{Focus Results}

For each environment (VR and reality), we check the extent to which the audience finds it easy to (1) focus on the presenter, (2) focus on the presentation content, and (3) shift focus between presenter and presentation content.

Looking at the results for (1) ($F(2,8)$=$0.63$, $p$>$0.1$) and (2) ($F(2,8)$<$0.01$, $p$>$0.1$), the difference in feeling when comparing the environments is small. P7(F) mentioned that the cursor on projected Chalktalk (displayed on a large monitor) is too small for her to follow and P1(M) emphasized that in VR the avatar's drawing hand fulfilled the role of a huge cursor for him, which helped him stay fully focused on the content. The result of frequency of shifting focus is ($M_{reality}$=$5.33$, $SD_{reality}$=$1.21$, $F(2,8)$=$11.36$, $p=0.02$). From that we can tell that the face-to-face design has a strong impact on how actively the audience shifts focus. In the interview, P5(F) described switching focus between presenter and content much more in the reality environment. Multiple participants mentioned that the face-to-face configuration helped them concentrate.

\subsection{One-on-one Results}

For the VR environment, we asked to what degree there was a feeling of one-on-one interaction, resulting in ($M$=$6.17$, $SD$=$0.75$). We also asked to what degree the audience member is aware of the other audience member: ($M_{VR}$=$1.5$, $SD_{VR}$=$0.84$, $F(2,8)$=$30.94$, $p$<$0.01$). In VR, although two audience members cannot see each other, we support group audio so both of them can speak to the presenter. All participants tried to speak to the presenter during the experiments in VR. The audience felt a sense of being given a one-on-one presentation when only the presenter was speaking (P2; P8, F), and they didn't encounter distractions in VR since the other audience member provides only audio feedback--no visual representation. This suggests that a one-on-one feeling helps the audience concentrate on the presentation.

\subsection{Retention}
The results of Q: ``which do you remember more in terms of the content" and Q: ``which do you remember more in terms of the platform" are both 83.3\% and 100\% think they can learn more in the VR platform. This suggests that for this specific topic, our system helps make content more memorable than projected Chalktalk.

\subsection{Enjoyment and Learning Results}

From the questionnaire, we see that there is no large difference in level of learning ($M_{VR}$=$6.67$, $M_{reality}$=$6.5$) and enjoyment ($M_{VR}$=$6.67$, $M_{reality}$=$5.67$) between VR and reality formats. P5(F) suggested that if the presentation platform itself is interesting, then it will help encourage the audience to be more focused and have a positive impact on the learning experience. That means that (1) she thinks that the VR setup is more intuitive and interesting to her and (2) that she can focus on the content more easily in VR than in other setups.

\subsection{Other Observations and Feedback}
\begin{itemize}
\item Deictic words were used in both environments, and based on the interview, we find that the audience has no great difficulty following them in either environment.
\item Audience members noticed that the VR environment has unlimited space. They found it helpful that the presenter could simply pan old content away without deleting it, making it possible to go back to previous material. P4(F) praised that previous content won’t disappear or be erased due to a lack of work-space and mentioned that this is important in that it helps students catch-up.  
\item Different participants had various opinions on the significance of eye contact. P6(M) thought that it was not very important,  and P5(F) preferred to look at the presenter at all times.
\item Some participants noted that the VR environment design, which features only the presenter, the audience participant, and content in a void environment,  helped them concentrate. In the regular presentation in reality, they felt more distant and more easily distracted.
\item Some participants (P5, F; P8, M) tried to move around and see the content from different views in VR. P5(F) said that she wanted to see the geometry from different angles so as to understand the matrix clearly.
\item P5(F) remarked that 10 minutes (around the same duration as the lesson in our experiment) was the longest time she could stay in VR.
\item 2 out of 8 prefer projected Chalktalk and the rest prefer our system.
\end{itemize}

\section{Guidelines}
\begin{itemize}
\item Presentation content oriented design

P5(F) mentioned the fact that she cannot use phones to help her stay focused on the content. She cannot take photos and write notes. A user-controlled screenshot-save feature is actually supported for both presenter and audience to solve this issue, but it is not included in the experiment. Still, P5(F) made the point that it is impossible for now to take notes in VR (if we don't integrate any other handwriting tool). So clearly the content of the presentation is an important factor. During the projected Chalktalk experiment, P7(F) commented that she does not want to do matrix calculation during the lecture. This also pointed out that if the lecture required the audience to do some precise handwriting work, our system wouldn't be a good choice. If the lesson requires mostly visualization and spatial cognition, then our system will be helpful.

\item Appearance of the avatar

P5(F) and P6(M) mentioned that because of the appearance of the presenter's avatar, they felt that they could not have eye contact with presenter. That means the appearance of the other members in the same environment is a significant factor in our ability to establish eye contact. For future research focused on the effectiveness of eye contact, a pre-experiment on the design of the avatar should be conducted first. After the appearance of the avatar becomes suitable for establishing eye contact (we will determine this through iterative testing and experiment), the formal experiment can continue. These pre-experiments will also impact findings on gaze discussion, as the avatar's apparent lack of eyes made some audience members such as (P7,F) find it difficult to see the presenter's gaze direction. Some additional cues such as a laser virtual pointer might also help with gaze awareness.

\item Multiple configuration for students

Most participants prefer the one-to-one experience, though. P5(F) said that she enjoys being in a group of other people during a presentation. She suggested that the system support an option for one audience member to see the others if desired. For example, if the students wish to have a group discussion, such as those in seminar-style classes, eye contact between them is also and important part of their interaction. To support different scenarios, supporting different preferences is almost certainly necessary.

\end{itemize}

\section{Conclusion}
We have proposed a framework that uses a data protocol to decouple the game engine and behavior server, and implemented an MR multi-user presentation system based on that framework. Then we conducted an experiment comparing projected Chalktalk with our system, using the topic of 4x4 matrix transformations to learn how participants experience these two systems. We showed observations from our comparison between projected Chalktalk and our MR platform, and guidelines for immersive presentation design.
It turns out that our system significantly decreases focus shift during presentation, which may help participants concentrate on content. Also, we learned that participants barely notice the existence of the other audience members, and received the experience as a one-on-one presentation with the presenter, which helped them to better concentrate on the presented material. We also confirmed some properties in MR are good for presentations, such as its lack of display space restrictions. Finally, we summarized design guidelines for MR presentations based on our results and feedback from user study interviews. The content and appearance of the participants, as well as customized support for different students were all found to be important design factors in this scenario.